\documentclass[sigconf, nonacm]{acmart}
\AtBeginDocument{%
  }

\setcopyright{acmlicensed}
\copyrightyear{2018}
\acmYear{2018}
\acmDOI{XXXXXXX.XXXXXXX}
\acmConference[Conference acronym 'XX]{Make sure to enter the correct
  conference title from your rights confirmation email}{June 03--05,
  2018}{Woodstock, NY}
\acmISBN{978-1-4503-XXXX-X/2018/06}

\usepackage{subcaption}
\usepackage[ruled,linesnumbered]{algorithm2e}

\begin{document}

\title{PrismStencil: Transforming Stencil Computations into Matrix Multiplications}

\author{Xianmeng Jiang}
\email{jiangxianmeng23z@ict.ac.cn}
\orcid{0000-0002-1664-7402}
\affiliation{%
  \institution{Institute of Computing Technology}
  \city{Haidian Qu}
  \state{Beijing Shi}
  \country{China}}

\author{Yuan Liang}
\correspondingauthor
\affiliation{%
  \institution{Institute of Computing Technology}
  \city{Haidian Qu}
  \state{Beijing Shi}
  \country{China}}
\email{yuanliang@ict.ac.cn}

\author{Zixuan Wang}
\affiliation{%
  \institution{Institute of Computing Technology}
  \city{Haidian Qu}
  \state{Beijing Shi}
  \country{China}}
\email{wangzixuan22@mails.ucas.ac.cn}

\author{Kun Li}
\affiliation{%
  \institution{Tsinghua University}
  \city{Haidian Qu}
  \state{Beijing Shi}
  \country{China}}
\email{likungw@gmail.com}

\author{Yunquan Zhang}
\affiliation{%
  \institution{Institute of Computing Technology}
  \city{Haidian Qu}
  \state{Beijing Shi}
  \country{China}}
\email{zyq@ict.ac.cn}

\renewcommand{\shortauthors}{Trovato et al.}

\begin{abstract}
Stencil computations are a fundamental class of workloads in scientific and engineering applications.
Although Tensor Cores offer new opportunities for stencil acceleration, existing tensorization methods suffer from substantial redundant computation and insufficient data reuse.
We propose PrismStencil, a stencil tensorization algorithm that introduces a novel mapping scheme between matrices and stencil grid points.
PrismStencil employs multidimensional vector mapping to accommodate the multidimensional data dependencies of stencil computations and a unified gather--scatter tensorization scheme to exploit the relative sizes of the matrix dimensions in Tensor Core operations, thereby reducing the redundant computation introduced by tensorization.
Moreover, it introduces a cross-product data-reuse mechanism that exploits the overlap among neighboring stencil updates to further enhance data reuse.
Across representative stencil kernels on an NVIDIA A100 GPU, PrismStencil achieves average speedups of 2.13x and 2.17x over state-of-the-art Tensor Core stencil methods with FP64 and TF32 precision, respectively.
\end{abstract}



\received{20 February 2007}
\received[revised]{12 March 2009}
\received[accepted]{5 June 2009}

\maketitle

\section{Introduction}
Stencil computation is one of the thirteen Berkeley motifs and represents a common class of nested-loop computations in scientific and engineering applications, dynamic programming, and image processing~\cite{SC15_Gysi,FPL20_Singh,glinski2001non,HeatTransfer_Acharya,ASPLOS15_Teja}.
A stencil computation operates on data arranged on a regular spatial grid with one or more dimensions and proceeds through a sequence of time steps.
At each time step, every grid point is updated using the values of a fixed set of nearby grid points from the previous time step.

A substantial body of work has been devoted to optimizing stencil computations, primarily through tiling~\cite{Grosser18, Bondhugula17, Grosser13, Yuan17, CGO20_Matsumura} and vectorization~\cite{Henretty13, Li22, Li21, Zhang25}.
Tiling changes the traversal order of grid points to improve temporal and spatial locality while exposing coarse-grained parallelism across tiles.
Vectorization processes multiple grid-point updates simultaneously using SIMD instructions, exposing fine-grained parallelism and increasing arithmetic throughput.

Tensor Cores offer substantially higher arithmetic throughput than standard SIMD units through their matrix multiply-and-accumulate (MMA) instructions.
Although originally designed for machine-learning workloads, they have increasingly been applied to scientific computing, including stencil computations.
In particular, recent studies~\cite{ICS22_Liu,SC24_Zhang,PPopp24_Chen,PPoPP26_Gu} have explored tensorizing stencil computations by organizing the weighted neighbor accumulations for multiple grid-point updates into matrix multiplications.

However, existing methods still fall short of fully exploiting this potential.
Prior work~\cite{gu2026needtensorcoresstencil} has shown that the matrix representations adopted by these Tensor Core stencil methods inevitably contain many zero elements, resulting in substantial redundant computation that offsets the computational advantage of Tensor Cores.
Furthermore, the matrix multiplications generated by stencil tensorization are typically small and therefore offer little opportunity for data reuse across MMA instructions through conventional matrix-multiplication blocking. Consequently, the available memory bandwidth is insufficient to keep Tensor Cores fully utilized.

To address these limitations, we propose PrismStencil, a stencil tensorization algorithm that introduces a novel mapping scheme between matrices and stencil grid points.
Unlike existing methods that impose a fixed correspondence between matrix-multiplication dimensions and spatial grid structures, PrismStencil maps each matrix-multiplication dimension across multiple spatial dimensions to better accommodate the multidimensional data dependencies of stencil computations, thereby constructing denser matrix representations.
PrismStencil also extends the tensorized computation mode to better exploit the relative sizes of the dimensions of an MMA instruction when organizing the stencil grid points involved in each matrix multiplication, further reducing the redundant computation introduced by tensorization.

Beyond the data reuse inherent in matrix multiplication itself, PrismStencil exploits the inherent overlap among neighboring stencil updates to expose additional reuse across matrix rows.
It further reorders data carefully across the matrix dimensions while preserving mathematical equivalence, extending such reuse to the granularity defined by MMA instruction shapes.
Together, these optimizations reduce repeated data movement and alleviate the memory-bandwidth bottleneck of tensorized stencil computation.

The main contributions of this work are as follows:
\begin{itemize}
    \item We propose a multidimensional vector mapping and a unified gather--scatter tensorization scheme that better accommodate the multidimensional structure of stencil computations and the dimensions of MMA instructions, substantially reducing the redundant computation introduced by tensorization.

    \item We develop a cross-product data-reuse mechanism that exploits the inherent overlap among neighboring stencil updates and enables data reuse at the granularity of MMA instruction shapes, thereby reducing repeated data movement.

    \item We implement PrismStencil for a diverse set of stencil kernels. Compared with state-of-the-art Tensor Core stencil methods, PrismStencil achieves average speedups of $2.13\times$ and $2.17\times$ for FP64 and TF32 computations, respectively.
\end{itemize}

\section{Background}
\label{sec:background}

\subsection{Stencils}

Stencil computations operate on a $d$-dimensional regular grid, where each grid point is iteratively updated based on a fixed neighborhood pattern bounded by stencil radius $r$, which specifies the maximum neighbor distance along each dimension.
Two common stencil patterns are \emph{star-shaped} stencils, which access only axis-aligned neighbors, and \emph{box-shaped} stencils, which access all neighbors within a hypercube of edge length $(2r+1)$ in $d$ dimensions, centered at the current grid point.
Algorithm~\ref{alg:2D Box Jacobi Stencil} illustrates a two-dimensional box-shaped stencil, where $u_{y,x}^{t}$ denotes the value at grid point $(y,x)$ at iteration $t$.
For each time step, we denote the input and output grid-point values by $a_{y,x}$ and $c_{y,x}$, respectively, so that the update can be written as $c_{y+r,x+r}=\sum_{i=0}^{2r}\sum_{j=0}^{2r} w_{i,j} \times a_{y+i,x+j}$.

\begin{algorithm}
\caption{2D Box Jacobi Stencil}\label{alg:2D Box Jacobi Stencil}
\For{$t \leftarrow 0$ \KwTo $T$}
{
    \For{$y \leftarrow 0$ \KwTo $NY$}
    {
        \For{$x \leftarrow 0$ \KwTo $NX$}
        {
            $u_{y+r,x+r}^{t+1}=\sum_{i=0}^{2r}\sum_{j=0}^{2r}w_{i,j}\times u_{y+i,x+j}^{t}$
        }
    }
}
\end{algorithm}

\subsection{Tensor Core}

Tensor Cores are specialized hardware units that accelerate matrix operations through matrix multiply-and-accumulate (MMA) instructions.
An MMA operation computes the product of two input matrices and accumulates the result into an output matrix:
\begin{equation}\label{eq:MMA}
D_{m\times n}=A_{m\times k}\times B_{k\times n}+C_{m\times n}.
\end{equation}
The matrix tiles involved in an MMA instruction, such as $A$, $B$, $C$, and $D$, are commonly referred to as \emph{fragments}.
The hardware-supported MMA shapes, denoted by $m\times n\times k$, are limited. For NVIDIA A100 GPUs, FP64 MMA supports only the $8\times 8\times 4$ shape, while TF32 MMA supports the $16\times 8\times 4$ and $16\times 8\times 8$ shapes~\cite{nvidia_ptx_instructions_9.1}.

\section{Challenge}
\label{sec:Challenge}

\subsection{Computational Inflation}
\label{sec:computational-inflation}

Stencil computations and MMA operations exhibit a natural mismatch in their computation patterns.
Therefore, using Tensor Cores to accelerate stencil computations often introduces redundant arithmetic.
To quantify this overhead, we define the computational inflation factor as the ratio of the floating-point operation count after tensorization to the intrinsic operation count of the stencil kernel.

Prior tensorization approaches can be classified into two categories according to their computation patterns.
The first category~\cite{ICS22_Liu,SC24_Zhang,PPoPP26_Gu} decomposes a high-dimensional stencil kernel into a sum of multiple one-dimensional stencil kernels, tensorizes each one-dimensional kernel separately, and accumulates the partial results.
For example, for a two-dimensional box-shaped stencil, these approaches compute a $y\times x$ output grid-point matrix in $(2r+1)$ steps.
Each step corresponds to a one-dimensional stencil kernel with radius $r$, where the input grid-point matrix has size $y\times (2r+x)$.
Since the input and output matrices must be aligned with the MMA fragment shape, the computational inflation factor is
\begin{equation}\label{eq:TCStencil-eff}
\frac{\text{align}(y,m) \times \text{align}(2r+x,k)\times \text{align}(x,n)}
{(2r+1)\times y\times x},
\end{equation}
where $\text{align}(x,y)=\lceil x/y\rceil \times y$.
The choices of $x$ and $y$ vary across different approaches.
TCStencil~\cite{ICS22_Liu} only discusses the case of $m=n=k$ and chooses $x=y=m-2r$.
LoRAStencil~\cite{SC24_Zhang} instead chooses $x=\text{align}(2r,n)$ and $y=\text{align}(2r,m)$ to avoid alignment overhead along the $m$ and $n$ dimensions.
It further exploits symmetric stencil coefficients for optimization.
For a fair comparison, however, we do not count the reduction enabled by stencil-kernel symmetry when analyzing the computation cost.

Different from the above approaches, ConvStencil~\cite{PPopp24_Chen} directly tensorizes high-dimensional stencil kernels.
Using its dual-tessellation algorithm, ConvStencil computes an $m\times (2r+2)$ output grid-point matrix in two steps.
In each step, the input grid-point matrix has size $m\times (2r+1)^2$.
Therefore, its computational inflation factor is
\begin{equation}\label{eq:ConvStencil-eff}
\frac{2\times \text{align}((2r+1)^2,k)\times \text{align}(2r+2,n)}
{(2r+1)^2\times (2r+2)} .
\end{equation}

Table~\ref{tab:TC-eff-cmp} compares the computational inflation factors of these approaches for common two-dimensional stencil kernels under the $8\times 8\times 4$ MMA shape.
For TCStencil, the computational inflation factor increases with the stencil radius $r$.
This is because its output matrix size, $(m-2r)^2$, decreases as the radius grows, leading to increasingly severe alignment overhead.
In contrast, the computational inflation factors of LoRAStencil and ConvStencil decrease as $r$ increases.
For small-radius kernels, these two approaches use kernel fusion~\cite{SC24_Zhang,PPopp24_Chen} to transform them into larger-radius kernels and thereby mitigate computational inflation.
However, this transformation also introduces extra computation~\cite{gu2026needtensorcoresstencil}.
For example, transforming three time steps of j2d9pt into one time step of j2d49pt introduces an additional computational inflation factor of $49/9/3=1.81$.
Consequently, after kernel fusion, the computational inflation factors of LoRAStencil and ConvStencil for j2d9pt become $4.16$ and $3.85$, respectively.
Compared with the cases without kernel fusion, these factors are not significantly reduced; for LoRAStencil, the factor even increases.
Moreover, even for large-radius kernels, Eq.~\eqref{eq:TCStencil-eff}, instantiated with $x=\text{align}(2r,n)$ and $y=\text{align}(2r,m)$, and Eq.~\eqref{eq:ConvStencil-eff} show that the computational inflation factors of LoRAStencil and ConvStencil asymptotically approach $2$ as $r$ grows.
This indicates that substantial redundant computation remains in these prior tensorization schemes.

\subsection{Memory Bound}

For modern GPUs, the data throughput of Tensor Cores is often much higher than the shared-memory bandwidth.
For example, on NVIDIA A100 GPUs, each Streaming Multiprocessor (SM) can issue one $16\times 8\times 4$ TF32 MMA instruction in one cycle on average, whereas loading the corresponding fragments $A$, $B$, and $C$ from shared memory requires 2, 1, and 4 cycles, respectively.
This imbalance makes data movement from shared memory a critical bottleneck for feeding Tensor Cores.

Conventional matrix multiplication alleviates this bottleneck through register-level blocking, which provides a general form of data reuse by sharing loaded matrix elements across multiple MMA instructions.
Such reuse becomes effective when the matrix dimensions $M$, $N$, and $K$ are sufficiently large to support substantial blocking.
However, as discussed in Section~\ref{sec:computational-inflation}, the matrix dimensions produced by stencil tensorization are typically comparable to the hardware-supported MMA shape $m\times n\times k$ and the stencil radius $r$.
Consequently, these matrix multiplications are too small for conventional register-level blocking to provide sufficient data reuse.
SPIDER~\cite{PPoPP26_Gu} retains the stencil-weight matrix in registers to avoid repeatedly loading the coefficients, but the input and output grid-point data in prior approaches~\cite{ICS22_Liu,SC24_Zhang,PPopp24_Chen,PPoPP26_Gu} still need to be repeatedly moved through shared memory.
This repeated movement prevents Tensor Cores from being fully utilized and makes the overall stencil computation bounded by shared-memory bandwidth.

\section{Algorithm}
\label{sec:Algorithm}

We first clarify the roles of the matrix dimensions.
For a stencil computation transformed into a matrix multiplication, let $A\in\mathbb{R}^{M\times K}$ denote the matrix formed by the input grid points, $B\in\mathbb{R}^{K\times N}$ the matrix formed by the stencil-kernel weights, and $C\in\mathbb{R}^{M\times N}$ the matrix formed by the output grid points.
The matrix multiplication can be viewed as $M$ vector--matrix products executed in parallel.
For each vector--matrix product, the input and output vectors correspond to $K$ input grid points and $N$ output grid points in the stencil grid, respectively.
For each input--output grid-point pair, the corresponding entry of $B$ is set to the stencil-kernel weight if the two grid points are neighbors under the stencil pattern, and is set to zero otherwise.
Therefore, the mapping of the input and output vectors to the stencil grid space determines the nonzero pattern of $B$ and directly affects computational inflation.
In contrast, the $M$ dimension indexes multiple such vector--matrix products.
It does not change the nonzero pattern of $B$ within each product, but determines how these products are placed in the grid space and thus affects data reuse across vector--matrix products.

PrismStencil redesigns the mapping between matrix dimensions and the stencil grid space according to stencil data dependences.
Section~\ref{sec:Tensorized Computation Schemes} focuses on the $K$ and $N$ dimensions and studies how to map the input and output vectors of each vector--matrix product to the stencil grid space, thereby increasing the fraction of effective input--output neighbor pairs and reducing computational inflation.
Section~\ref{sec:Cross-Product Data Reuse} then considers the $M$ dimension and the full matrix multiplication.
By arranging vector--matrix products to expose cross-product data reuse, we reduce repeated shared-memory traffic and alleviate the memory-bandwidth bottleneck.

\subsection{Tensorized Computation Schemes}
\label{sec:Tensorized Computation Schemes}

\begin{figure}[t]
 \centering
 \begin{subfigure}{0pt}
    \centering
    \includegraphics[width=1\linewidth]{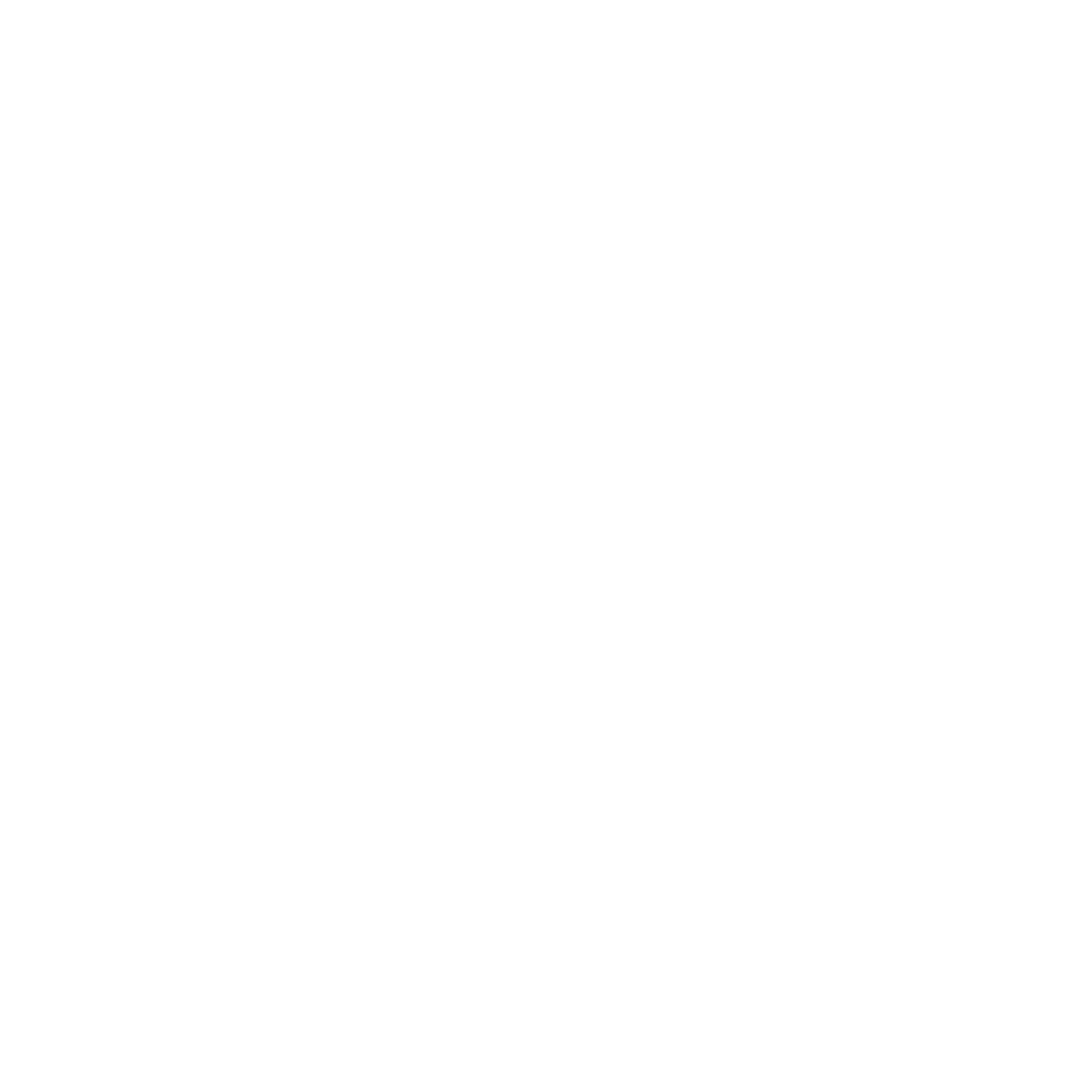}
  \end{subfigure}
  \begin{subfigure}{0.47\textwidth}
    \centering
    \includegraphics[width=1\linewidth]{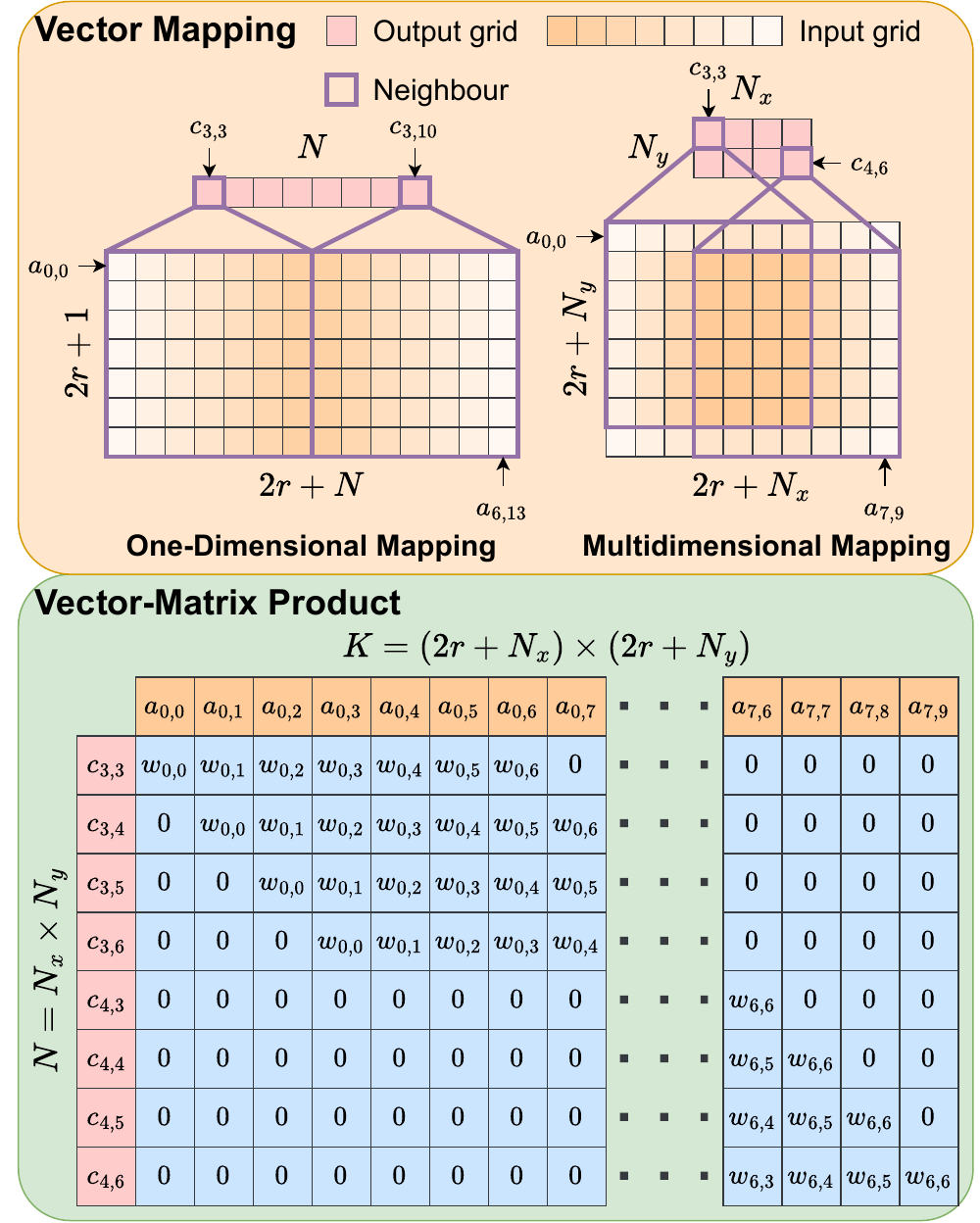}
  \end{subfigure}
\caption{Comparison between the one-dimensional mapping used in prior work and the multidimensional mapping in PrismStencil. In the upper panel, each input grid point is shaded according to the number of its neighboring grid points in the output vector, with darker colors indicating more neighbors. The lower panel illustrates the corresponding construction of the stencil-weight matrix $B$.}
\label{img:TCGAStencil}
\end{figure}

\subsubsection{Multidimensional Vector Mapping}
\label{sec:Multidimensional Mapping}

As in prior tensorization approaches~\cite{ICS22_Liu,SC24_Zhang,PPopp24_Chen,PPoPP26_Gu}, we begin with a vector--matrix product that performs a single time-step update for $N$ output grid points.
The length-$K$ input vector contains the union of the neighboring input grid points required by these outputs.

The computational inflation of this vector--matrix product depends on the neighborhood overlap induced by the output-vector mapping.
Larger overlap yields a more compact input-neighbor union and therefore a higher fraction of valid input--output pairs.
Prior approaches~\cite{ICS22_Liu,SC24_Zhang,PPopp24_Chen,PPoPP26_Gu} map the output vector to an elongated one-dimensional region.
Under such a mapping, substantial neighborhood sharing occurs only among nearby points along that dimension, while distant points, especially those near opposite endpoints, share few or no neighboring input grid points.
This mapping is inefficient for multidimensional stencils, whose data dependences extend across multiple dimensions.

As illustrated in Figure~\ref{img:TCGAStencil}, PrismStencil introduces a multidimensional vector mapping that distributes the output vector across multiple stencil dimensions, forming a compact region in the stencil grid space.
This mapping allows output grid points to share neighboring input grid points along multiple dimensions, reducing the size of the input-neighbor union and hence the number of redundant input--output pairs.

Formally, we characterize the above output-vector mapping by considering a $d$-dimensional stencil with radius $r$.
Suppose the length-$N$ output vector is mapped to a multidimensional region that contains $N_i$ output grid points along the $i$-th dimension, where
$N=\prod_{i=1}^{d} N_i$.
To update these output grid points, the length-$K$ input vector is mapped to the union of their neighboring input grid points, which contains
$K=\prod_{i=1}^{d}(2r+N_i)$
grid points.
Since each output grid point has $(2r+1)^d$ neighbors, the computational inflation factor is given by
\begin{equation}\label{eq:TCGAStencil-eff}
\frac{\text{align}(K,k)\times \text{align}(N,n)}
{(2r+1)^d\times N},
\end{equation}
where $\text{align}(x,y)=\lceil x/y\rceil \times y$.
Here, the alignment accounts for the fixed fragment sizes of MMA instructions: the $K$ and $N$ dimensions must be padded to multiples of the hardware-supported fragment dimensions $k$ and $n$, respectively.
For a fixed $N$, the values of $N_i$ should be kept as balanced as possible to minimize $K$, thereby minimizing the computational inflation factor in Eq.~\eqref{eq:TCGAStencil-eff}.

The remaining question is how to choose the length $N$ of the output vector.
In practice, we set $N=n$ to match the hardware-supported size of the $N$ dimension.
Increasing $N$ weakens the neighborhood overlap within the output vector and increases redundant computation, whereas decreasing $N$ incurs significant hardware-alignment overhead.
A smaller $N$ can be preferable only in rare factorization cases.
For example, if $n=17$ in a 2D stencil, mapping the output vector to a $4\times4$ region may be better than mapping it to a $1\times17$ region, despite leaving one output position unused.
However, hardware-supported MMA dimensions are typically powers of two, so such cases do not arise in the MMA shapes considered in this work.

As shown in Table~\ref{tab:TC-eff-cmp}, the multidimensional vector mapping reduces computational inflation by $33.3\%$, $20.0\%$, and $23.1\%$ for j2d9pt, j2d25pt, and j2d49pt, respectively, compared with the best-performing prior approach.

\subsubsection{Unified Gather--Scatter Tensorization Scheme}
\label{sec:Mode Extension}

The tensorization design described above corresponds to a gather-style scheme for stencil computation.
While the gather scheme lets selected output grid points accumulate weighted values from neighboring input grid points, the scatter scheme reverses this direction by letting selected input grid points contribute their weighted values to neighboring output grid points.
In prior work~\cite{ICS22_Liu,SC24_Zhang,PPopp24_Chen,PPoPP26_Gu}, the gather computation direction remains implicit, leaving the symmetric scatter-style alternative unexplored.
In scalar or vectorized stencil implementations, gather and scatter formulations perform the same amount of computation.
In tensorized computation, however, this symmetry no longer holds because the hardware-supported sizes of the $K$ and $N$ dimensions, i.e., $k$ and $n$, are not necessarily equal, creating additional opportunities to reduce redundant computation.

Relative to the gather-style scheme, a scatter-style scheme can be constructed by symmetrically reversing the computation direction.
Specifically, suppose the length-$K$ input vector is mapped to a multidimensional region that contains $K_i$ input grid points along the $i$-th dimension, where $K=\prod_{i=1}^{d}K_i$.
The output vector is then mapped to the union of the neighboring output grid points of these inputs, a set of size $N=\prod_{i=1}^{d}(2r+K_i)$.
By the same reasoning, we keep the values of $K_i$ as balanced as possible and set $K=k$ to minimize the computational inflation factor.
When $k<n$, the input-vector mapping in the scatter-style scheme covers fewer grid points than the output-vector mapping in the gather-style scheme, causing their neighboring point sets to overlap more extensively and thus reducing computational inflation.

We further generalize the gather and scatter schemes into a unified gather--scatter tensorization scheme.
For clarity, assume that the first $\omega$ dimensions follow the gather style, while the remaining dimensions follow the scatter style.
Along each gather-style dimension $1\leq i\leq\omega$, the output vector spans $N_i$ output grid points.
The input vector must therefore cover the corresponding neighboring input range, which contains $2r+N_i$ input grid points, and the number of effective input--output neighbor pairs along this dimension is $(2r+1)N_i$.
Similarly, along each scatter-style dimension $\omega<i\leq d$, the input vector spans $K_i$ input grid points.
The output vector must then cover the corresponding neighboring output range, which contains $2r+K_i$ output grid points, and the number of effective input--output neighbor pairs along this dimension is $(2r+1)K_i$.
Thus, the input and output vectors in the resulting vector--matrix product have lengths
$K=\prod_{i=1}^{\omega}(2r+N_i)\times \prod_{i=\omega+1}^{d}K_i$
and
$N=\prod_{i=1}^{\omega}N_i\times \prod_{i=\omega+1}^{d}(2r+K_i)$,
respectively.
The corresponding computational inflation factor is then given by
\begin{equation}\label{eq:TCSGStencil-eff}
\frac{\text{align}(K,k)\times \text{align}(N,n)}
{(2r+1)^d\times \prod_{i=1}^{\omega}N_i\times \prod_{i=\omega+1}^{d}K_i}.
\end{equation}

\begin{figure}[t]
\centerline{\includegraphics[width=0.47\textwidth]{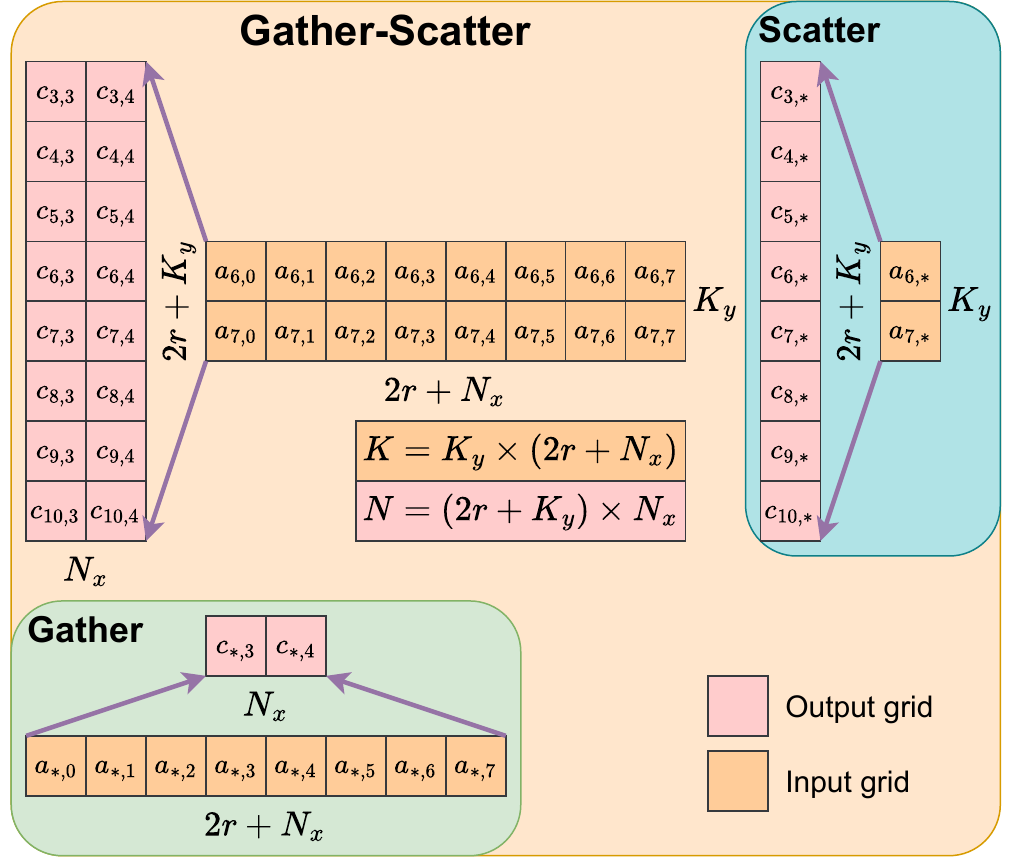}}
\caption{A mixed gather--scatter tensorization example for the j2d49pt stencil kernel with $N_x=2$ and $K_y=2$.}
\label{img:TCSGStencil}
\end{figure}

When $0<\omega<d$, the scheme corresponds to a mixed gather--scatter form.
By applying gather and scatter choices along different stencil dimensions, this form makes $K$ and $N$ more balanced, creating opportunities to use smaller values of $N_i$ and $K_i$ while still avoiding excessive alignment overhead.
For example, for the j2d49pt kernel with $n=k=8$, the gather and scatter schemes map the output or input vector to a $2\times4$ region, both resulting in a computational inflation factor of $1.63$.
In contrast, the mixed scheme can choose $N_x=2$ and $K_y=2$, which keeps both $K$ and $N$ aligned while further increasing the fraction of effective neighbor pairs.
As a result, the computational inflation factor is reduced to $1.31$.
Figure~\ref{img:TCSGStencil} illustrates the computation process under this configuration.

Finally, given a stencil radius $r$ and an MMA shape $m \times n \times k$, we use Eq.~\eqref{eq:TCSGStencil-eff} to select the computation scheme with the lowest computational inflation factor.
As shown in Table~\ref{tab:TC-eff-cmp}, building on the multidimensional vector mapping, the unified gather--scatter scheme further reduces the computational inflation factor by $33.3\%$, $16.7\%$, and $19.6\%$ for j2d9pt, j2d25pt, and j2d49pt, respectively.
More importantly, Table~\ref{tab:TC-eff-cmp} shows that our scheme is the first to reduce tensorization-induced computational inflation below $2.0$ on a practical hardware-supported MMA shape.
This indicates that our tensorized algorithm can still provide effective performance benefits over vectorized implementations even on platforms where MMA hardware provides only twice the compute throughput of SIMD hardware.

\begin{table}[t]
\caption{Theoretical computational inflation factors of different tensorization methods for different stencil kernels under the $8\times8\times4$ MMA shape.}
\begin{center}
\begin{tabular}{*{4}{c}}
\toprule
Method                          & j2d9pt & j2d25pt & j2d49pt\\
\midrule
TCStencil~\cite{ICS22_Liu}      & 4.74   & 6.40    & 18.29\\
LoRAStencil~\cite{SC24_Zhang}   & 4.00   & 2.40    & 2.29\\
ConvStencil~\cite{PPopp24_Chen} & 5.33   & 2.98    & 2.12\\
Our (Section~\ref{sec:Multidimensional Mapping})                     & 2.67   & 1.92    & 1.63\\
Our (Section~\ref{sec:Mode Extension})                    & 1.78   & 1.60    & 1.31\\
\bottomrule
\label{tab:TC-eff-cmp}
\end{tabular}
\end{center}
\end{table}

\subsection{Cross-Product Data Reuse}
\label{sec:Cross-Product Data Reuse}

\begin{figure}[t]
\centerline{\includegraphics[width=0.47\textwidth]{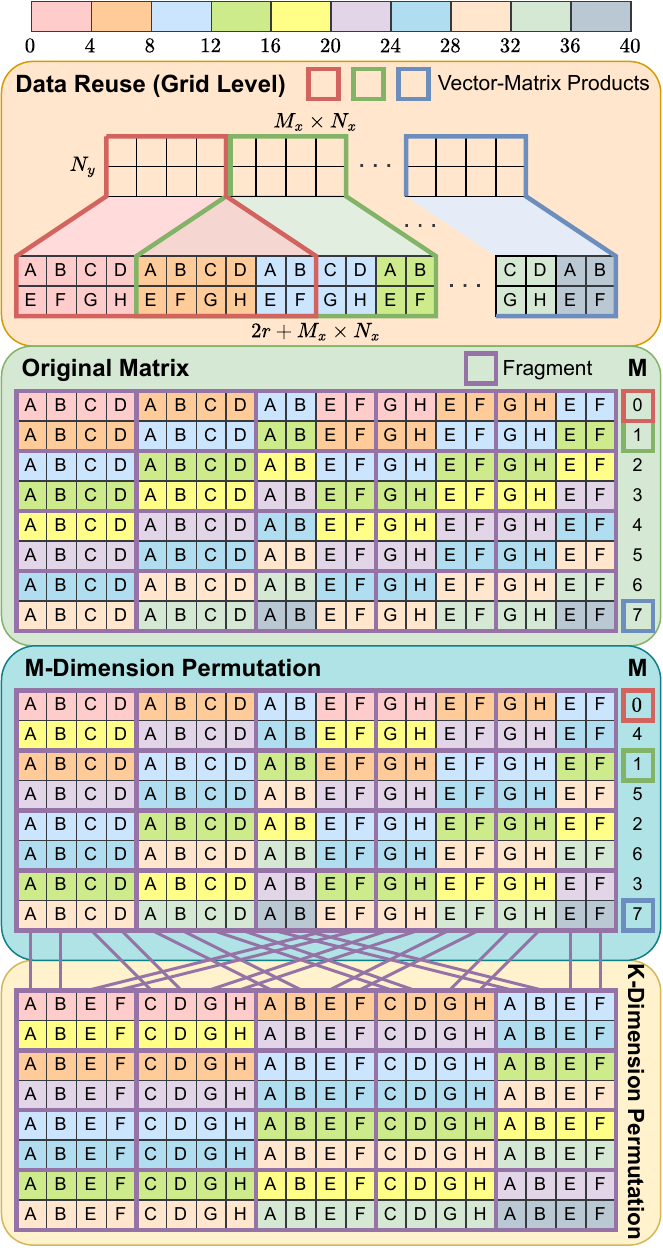}}
\caption{Overview of the matrix-reordering algorithm, using the gather-style tensorization of the j2d49pt kernel with an MMA shape of $m=2$, $n=8$, and $k=4$ as an example.
The letter and color of each grid point jointly encode its coordinates in the grid space.
For clarity, only the first two grid rows represented by the input matrix are shown.}
\label{img:Matrix reorder}
\end{figure}

We next consider the $M$ dimension and the full matrix multiplication, focusing on data reuse across different vector--matrix products.
As illustrated in Figure~\ref{img:Matrix reorder}, unlike conventional matrix multiplication, the stencil neighborhood pattern can cause vectors from different matrix rows to overlap in the grid space.
Such overlap occurs among input vectors in the gather scheme and among output vectors in the scatter scheme, enabling additional cross-row data reuse.

Similar to multidimensional vector mapping, we distribute the $M$ dimension across multiple stencil dimensions to exploit the data reuse enabled by multidimensional stencil dependences.
Let $M_i$ denote the number of adjacent vector--matrix products arranged along the $i$-th dimension, such that $M=\prod_{i=1}^{d}M_i$.
For a gather-style dimension, each vector--matrix product individually requires $2r+N_i$ input grid points along that dimension.
However, $M_i$ adjacent products jointly require only $2r+M_i\times N_i$ input grid points, reducing the amortized number per product to $2r/M_i+N_i$.
Similarly, for a scatter-style dimension, arranging $M_i$ adjacent products reduces the amortized number of output grid points written by each product from $2r+K_i$ to $2r/M_i+K_i$.
Thus, the amortized numbers of input and output grid points accessed by each vector--matrix product are $\prod_{i=1}^{\omega}\left(2r/M_i+N_i\right)\times\prod_{i=\omega+1}^{d}K_i$ and $\prod_{i=1}^{\omega}N_i\times\prod_{i=\omega+1}^{d}\left(2r/M_i+K_i\right)$, respectively.

\begin{figure}[t]
\centerline{\includegraphics[width=0.47\textwidth]{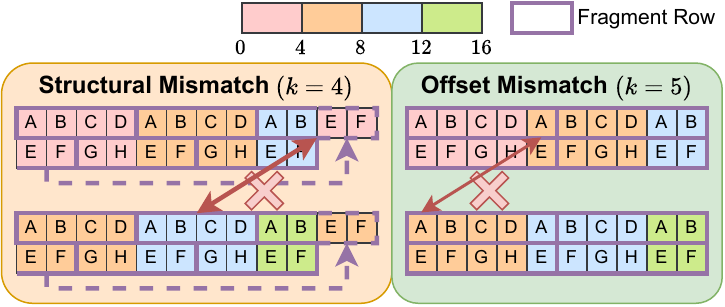}}
\caption{Illustration of structural and offset mismatches that prevent data reuse at the fragment-row granularity.}
\label{img:Mismatch}
\end{figure}

At the global-to-shared-memory level, exploiting this data reuse is straightforward.
Before computation, we load the complete set of required input grid points from global memory into shared memory, and after computation, we write the corresponding output grid points back to global memory.
For MMA instructions, however, data can be reused only at the fragment granularity rather than at the granularity of individual grid points.
As illustrated in Figure~\ref{img:Matrix reorder}, if the $\prod_{i=1}^{d}M_i$ vector--matrix products are mapped to the $M$ dimension in lexicographic order and the input and output grid points within each product are likewise mapped to the $K$ and $N$ dimensions in lexicographic order, the reusable grid points generally do not form identical layouts across the corresponding fragments, preventing direct fragment-level reuse.

PrismStencil addresses this mismatch by reordering the $M$, $N$, and $K$ dimensions, as illustrated in Figure~\ref{img:Matrix reorder}.
These transformations preserve mathematical equivalence because the corresponding rows and columns of the matrices representing input grid points, output grid points, and stencil weights are permuted consistently.
In practice, no physical data rearrangement is required; the reordering is realized simply by changing the grid-point coordinates from which each fragment is loaded and to which it is stored.

We first reorder the $M$ dimension so that reusable grid points are aligned to the same row position within different fragments whenever possible, where $m$ denotes the fragment size along the $M$ dimension, i.e., the number of rows in each fragment associated with grid-point data.
Consider a vector--matrix product with lexicographic index $(\alpha_{d-1},\dots,\alpha_1,\alpha_0)$, where $\alpha_i$ denotes its coordinate among the $M_i$ products along the $i$-th stencil dimension.
We map this product to row $\alpha$ of the $M$ dimension according to
\begin{equation}\label{eq:M-dim}
\alpha=
\sum_{i=1}^{d-1}\alpha_i\prod_{j=0}^{i-1}M_j
+\left(\alpha_0\bmod\frac{M_0}{m}\right)\times m
+\left\lfloor\frac{\alpha_0}{M_0/m}\right\rfloor.
\end{equation}
The last two terms interleave the products along the spatially contiguous dimension, indexed by $0$, across different fragments: $\alpha_0\bmod(M_0/m)$ determines the fragment, while $\lfloor\alpha_0/(M_0/m)\rfloor$ determines the intra-fragment row.
Consequently, adjacent products along this dimension are placed in the same row of different fragments, prioritizing them for reuse because they share substantially more grid points than nonadjacent products, while the remaining dimensions retain lexicographic ordering through the first term.
For simplicity, we restrict $M_0$ to multiples of $m$ throughout this work; this restriction is not fundamental but avoids more complicated mappings and implementation logic.

After reordering the $M$ dimension, some fragments still cannot be directly reused because of two types of layout mismatch, as illustrated in Figure~\ref{img:Mismatch}.
The first is a \emph{structural mismatch}.
In the example, the input region has length $N_x+2r$ along the $x$ dimension, which is not divisible by the row length $k$ of fragment $A$.
Consequently, lexicographic mapping causes a fragment row to cross the boundary of the $x$ dimension, and corresponding grid points in adjacent vector--matrix products are grouped with different neighboring points.
The second is an \emph{offset mismatch}.
Because the offset $N_x$ between adjacent vector--matrix products is not divisible by $k$, corresponding grid points are mapped to different columns within their fragments.
Moreover, $k$ is fixed by the hardware and cannot be adjusted directly to eliminate these mismatches.

To address this problem, we further map the $k$ and $n$ dimensions of an MMA fragment across multiple stencil dimensions, such that $k=\prod_{i=0}^{d-1}k_i$ and $n=\prod_{i=0}^{d-1}n_i$.
Equivalently, we reorder the $K$ and $N$ dimensions in a block-major layout: grid points are first grouped into multidimensional blocks of shape $(k_{d-1},\dots,k_0)$ or $(n_{d-1},\dots,n_0)$, and are then ordered lexicographically both across blocks and within each block.
To eliminate both structural and offset mismatches, each gather-style dimension must satisfy $k_i\mid(N_i+2r)$ and $k_i\mid N_i$, while each scatter-style dimension must satisfy $n_i\mid(K_i+2r)$ and $n_i\mid K_i$.
For example, Figure~\ref{img:Matrix reorder} chooses $k_y=2$ and $k_x=2$, allowing nearly all fragments containing reusable grid points to be directly reused.

The above constraints do not always admit valid choices of $k_i$ and $n_i$.
We can further relax them by exploiting the register-level operand layout of MMA instructions.
For example, on NVIDIA A100 GPUs, for the $m=8$, $n=8$, and $k=4$ MMA shape, each thread holds fragment $C$ in two registers.
The minimum granularity for reusing its data is therefore smaller than the entire fragment, allowing us to use an effective register-level size of $n_{R}=4$ in place of $n$ and relax the constraint $\prod_{i=0}^{d-1}n_i=n$ accordingly.
With such register-level relaxation, valid choices of $k_i$ and $n_i$ exist on all hardware platforms evaluated in this work.
On a platform where no valid factorization can be found, the algorithm can fall back to the $M$-dimension reordering alone and still exploit partial data reuse at the fragment granularity.

\section{Implementation}

On NVIDIA A100 GPUs, we assign each $M\times N\times K$ matrix multiplication described in Section~\ref{sec:Algorithm} to a single warp.
This design follows the warp-level programming interface of MMA instructions and ensures that fragment-level data reuse remains within the same warp, as fragments cannot be directly shared across warps.
This ownership is particularly important for scatter-style schemes.
The overlap of output grid points among different vector--matrix products creates opportunities for data reuse but also introduces potential write conflicts.
By assigning the entire matrix multiplication to one warp, all MMA instructions that update the same output fragment are issued within a single instruction stream, allowing the compiler and hardware to coordinate their execution without write conflicts.

\begin{figure}[t]
\centerline{\includegraphics[width=0.47\textwidth]{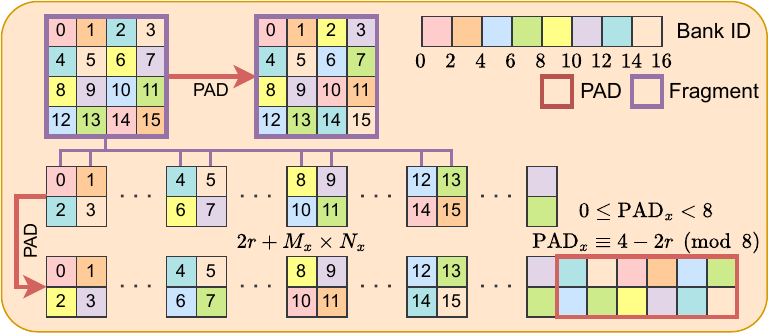}}
\caption{Illustration of eliminating shared-memory bank conflicts through padding. The number on each grid point denotes the thread that loads it, while the color indicates its shared-memory bank assignment. For simplicity, the example assumes a 16-thread warp and 16 shared-memory banks. With FP64 data, each grid point occupies two banks, and each fragment is loaded in two phases by the first and second groups of eight threads. A load is conflict-free when the grid points accessed by the eight threads in each phase have distinct colors.}
\label{img:smem bank}
\end{figure}

In addition, the matrix reordering introduced in Section~\ref{sec:Cross-Product Data Reuse} improves fragment-level data reuse but also makes the grid-space distribution of elements within each fragment more irregular, potentially introducing shared-memory bank conflicts.
For fragment $B$, the stencil-weight matrix is small and remains fixed throughout the computation.
We therefore preprocess its layout so that loading fragment $B$ follows a contiguous access pattern.
For fragments $A$ and $C$, as illustrated in Figure~\ref{img:smem bank}, we add appropriate padding to their shared-memory layouts to eliminate bank conflicts.

Finally, we adopt N.5D blocking~\cite{IEEE18_Rawat,CGO20_Matsumura,ICS23_Zhang} to organize computation across warps and thread blocks, thereby exposing parallelism at both levels. This blocking scheme also improves data reuse between global and shared memory, reducing the associated data-transfer overhead.

\section{Experimental Results}\label{sec:exp}

\subsection{Setup}\label{sec:setup}

\subsubsection{Platform}
All experiments were conducted on an NVIDIA A100 80GB SXM GPU.
CUDA 12.9 was used with the \texttt{-O3 --gpu-architecture=native} compiler options enabled.

\subsubsection{Benchmark}
As summarized in Table~\ref{tab:Stencil benchmark}, we selected eight representative stencil kernels following prior work~\cite{CGO20_Matsumura,PPopp24_Chen}. All benchmarks were evaluated in both TF32 and FP64 precision. To ensure fair comparison across different implementations, all stencil kernels employed non-symmetric coefficients.
\begin{table}[h]
\caption{Stencil benchmark}
\begin{center}
\begin{tabular}{*{3}{c}|*{3}{c}}
\toprule
Kernel      & RAD    & $FS^{\mathrm{a}}$             & Kernel  & RAD    & $FS^{\mathrm{a}}$ \\
\hline
j2d5pt      & $1$    & $10$            & j2d9pt   & $1$    & $18$            \\
j2d25pt     & $2$    & $50$            & j2d49pt  & $3$    & $98$              \\
j2d81pt     & $4$    & $162$           & j3d7pt   & $1$    & $14$            \\
j3d27pt     & $1$    & $54$            & j3d125pt & $2$    & $250$           \\
\bottomrule
\multicolumn{6}{l}{$^{\mathrm{a}}$ $FS$ denotes FLOPs per stencil.}
\label{tab:Stencil benchmark}
\end{tabular}
\end{center}
\end{table}

\subsubsection{State-of-the-Art Baselines}

PrismStencil was compared against state-of-the-art CUDA Core implementations, including AN5D~\cite{CGO20_Matsumura} and EBISU~\cite{ICS23_Zhang}, and Tensor Core-based implementations, including TCStencil~\cite{ICS22_Liu}, ConvStencil~\cite{PPopp24_Chen}, LoRAStencil~\cite{SC24_Zhang}, and SPIDER~\cite{PPoPP26_Gu}.
AN5D and EBISU provide both FP64 and FP32 implementations and are used as the CUDA Core baselines in the corresponding experiments.
ConvStencil and LoRAStencil provide only FP64 implementations and are therefore excluded from the TF32 comparison.
TCStencil and SPIDER provide only FP16 implementations.
Their TF32 performance is estimated as half of their measured FP16 performance, because both Tensor Core throughput and memory-system throughput for TF32 are half of those for FP16 on the evaluated GPU.

Strictly speaking, the FP32 performance of CUDA Core implementations is not directly comparable with the TF32 performance of Tensor Core implementations because the two formats provide different numerical precision.
The comparison is included primarily to evaluate the relative performance advantage of Tensor Core-based stencil computation in a setting where Tensor Cores provide approximately eight times the arithmetic throughput of CUDA Cores.

\subsubsection{Problem Size}\label{sec:Problem Size}

Problem sizes were selected following prior work~\cite{ICS23_Zhang}.
Slightly different sizes were used for different algorithms to avoid incomplete tiling and ensure a fair comparison, while keeping the spatial dimensions as similar as possible.
Representative configurations are reported in Table~\ref{tab:Problem Size A100 part}, and the complete set is provided in the Supplemental Material.

\subsubsection{Metrics}

Performance is reported in TFLOPS and calculated as
\begin{equation}
\label{eq:gstencil}
\text{TFLOPS}
=\frac{FS\times \text{Stencil/s}}{10^{12}}
=\frac{FS\times T\times \prod_{i=1}^{N}\text{Dim}_i}
{t\times 10^{12}}.
\end{equation}
Here, $FS$ is given in Table~\ref{tab:Stencil benchmark}, $T$ denotes the size of the temporal dimension, $N$ is the number of spatial dimensions, and $\text{Dim}_i$ denotes the size of the $i$-th spatial dimension.
The numerator therefore counts only the floating-point operations required by the original stencil algorithm, and the reported TFLOPS represents the effective algorithmic throughput rather than the actual floating-point throughput executed by the hardware.
The variable $t$ denotes the kernel execution time in seconds and thus captures all implementation overheads, including additional computation caused by computational inflation.
After appropriate warm-up, each kernel was measured ten times, and the peak performance was reported.

\begin{table}[t]
\caption{Problem size configuration for FP64 on NVIDIA A100}
\begin{center}
\begin{tabular}{*{3}{c}}
\toprule
Method       & j2d9pt & j3d27pt \\
\midrule
ConvStencil  & $1000\times 9216^2$  & $1000\times 2560\times 288\times 384$ \\
LoRAStencil  & $1000\times 9216^2$  & $1000\times 1280\times 480\times 576$ \\
AN5D         & $1000\times 8192^2$  & $1000\times 640\times 640\times 640$ \\
EBISU        & $1000\times 8784^2$  & $1000\times 2560\times 288\times 384$ \\
PrismStencil & $1000\times 9648^2$  & $1000\times 2560\times 270\times 368$ \\
\bottomrule
\label{tab:Problem Size A100 part}
\end{tabular}
\end{center}
\end{table}

\subsection{Sensitivity to the $M$ Dimension}

\begin{figure}[t]
 \centering
 \includegraphics[width=0.96\linewidth]{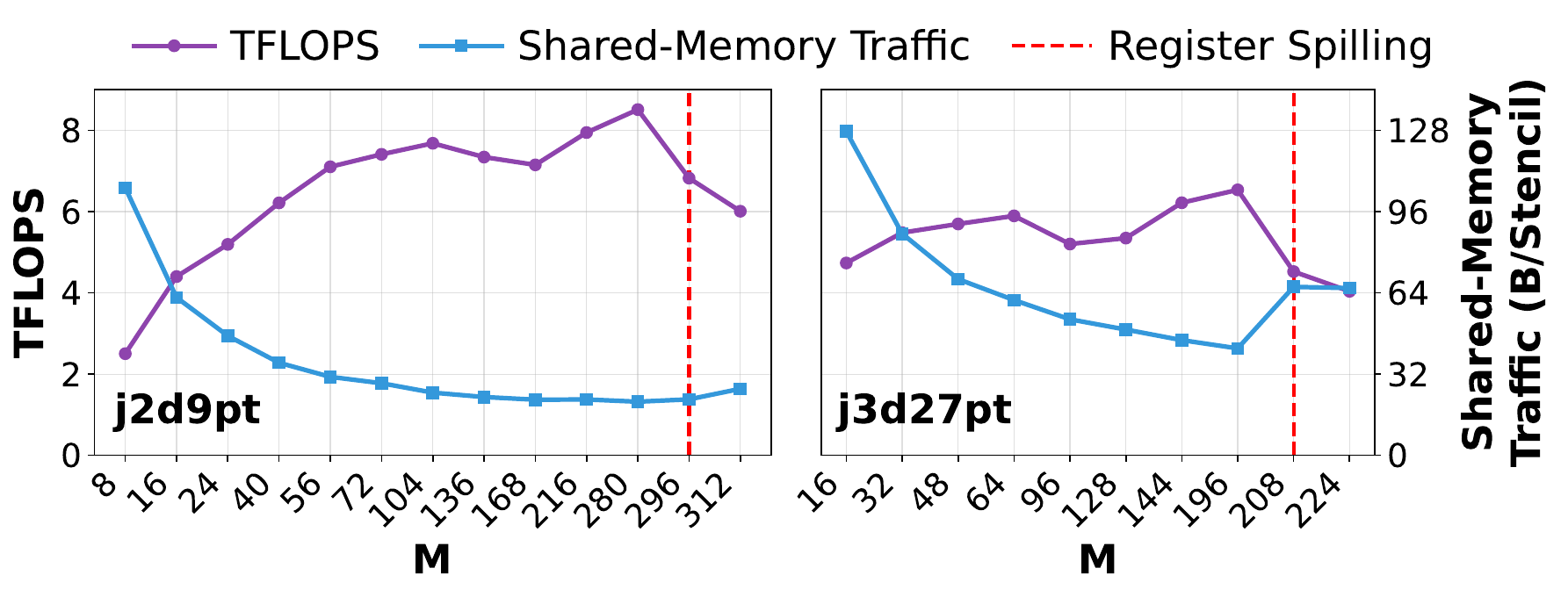}
\caption{Sensitivity of PrismStencil to the $M$ dimension under FP64 precision. B/Stencil denotes the amortized shared-memory traffic per grid-point update in one time step.}
\label{fig:sensitivity}
\end{figure}

Figure~\ref{fig:sensitivity} shows the performance and shared-memory traffic of PrismStencil as the size of the $M$ dimension varies.
As $M$ increases, the shared-memory traffic steadily decreases until register spilling occurs.
This reduction gradually approaches the minimum grid-point traffic required by each stencil update: one read and one write per grid point, corresponding to $16$ bytes per stencil update for FP64 data.

The performance of PrismStencil also exhibits an overall increasing trend with $M$, primarily benefiting from the reduced shared-memory traffic, and reaches its maximum immediately before register spilling.
Although increasing $M$ raises per-thread register usage and consequently lowers thread occupancy, it also exposes more independent operations within each matrix multiplication.
The resulting increase in instruction-level parallelism compensates for the reduction in thread-level parallelism, leaving the overall parallelism largely unaffected.
Based on these results, all subsequent experiments select the largest value of $M$ that does not cause register spilling.

\subsection{State-of-the-Art Comparison}

\begin{figure*}[t]
  \centering
  \begin{subfigure}{0.88\textwidth}
    \centering
    \includegraphics[width=1\textwidth]{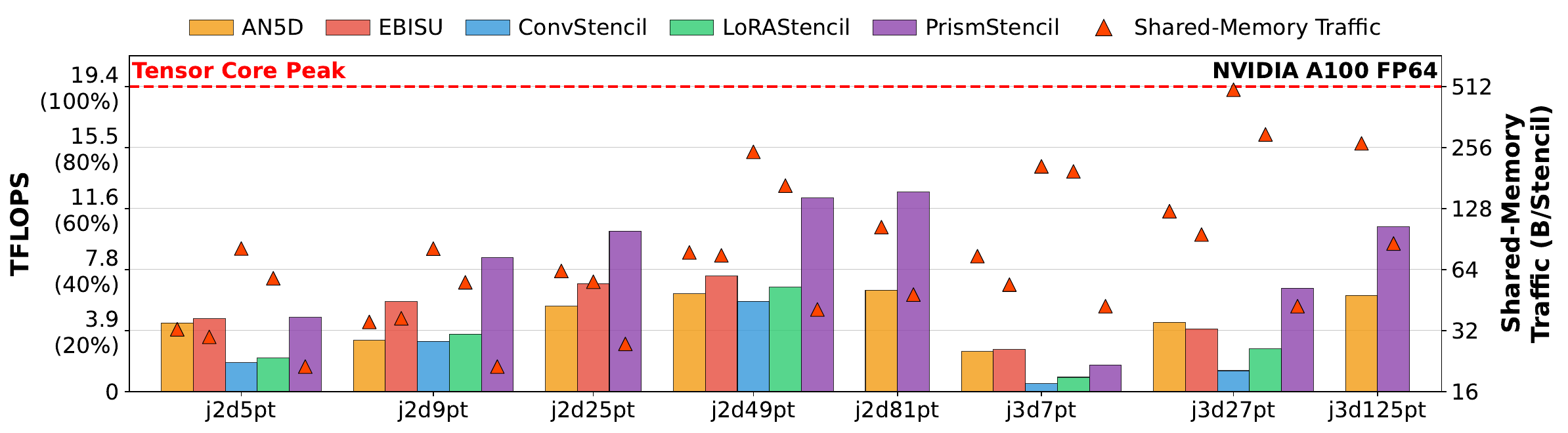}
    \label{img:perf-A100-fp64}
  \end{subfigure}
  \hfill
  \begin{subfigure}{0.88\textwidth}
    \centering
    \centerline{\includegraphics[width=1\textwidth]{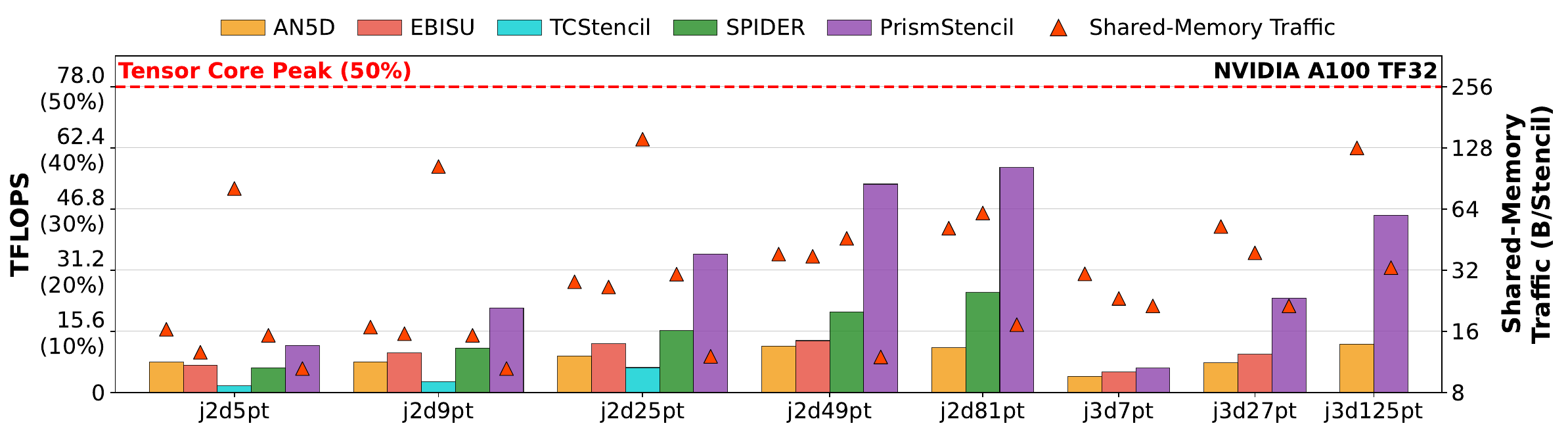}}
    \label{img:perf-A100-tf32}
  \end{subfigure}
\caption{Comparison of PrismStencil with state-of-the-art approaches in terms of performance and shared-memory traffic. Bars show performance (higher is better), while upward-pointing triangles show shared-memory traffic in B/Stencil (lower is better). For TCStencil and SPIDER, the shared-memory traffic is multiplied by two because their FP16 data elements are half the width of TF32 elements.}
\label{img:perf}
\end{figure*}

Figure~\ref{img:perf} compares PrismStencil with state-of-the-art approaches in terms of performance and shared-memory traffic under FP64 and TF32 precision.

On average, PrismStencil achieves $3.63\times$ higher performance with TF32 than with FP64.
This improvement is lower than the $8\times$ difference in peak Tensor Core throughput between the two precisions.
This gap arises because the shared-memory system processes TF32 elements at only twice the rate of FP64 elements, causing the arithmetic-to-memory throughput ratio to increase by a factor of four under TF32.
Consequently, supplying data to Tensor Cores becomes relatively more demanding, making it more difficult for TF32 kernels to approach their peak arithmetic throughput.

Compared with state-of-the-art Tensor Core approaches, PrismStencil consistently demonstrates substantial performance advantages and lower shared-memory traffic.
Under FP64 precision, PrismStencil achieves an average speedup of $2.13\times$ over LoRAStencil while reducing shared-memory traffic by $72.9\%$ on average.
Under TF32 precision, it achieves an average speedup of $2.17\times$ over SPIDER and reduces shared-memory traffic by $53.8\%$ on average.

Compared with state-of-the-art CUDA Core approaches, PrismStencil provides limited speedups for the star-shaped j2d5pt and j3d7pt stencils.
These kernels have relatively low arithmetic intensity and are therefore more severely constrained by memory traffic.
Moreover, adjacent grid points in star-shaped stencils share fewer neighbors, resulting in less data reuse and greater computational inflation for Tensor Core-based computation.
For box-shaped stencils, PrismStencil achieves average speedups of $1.94\times$ and $3.81\times$ over AN5D under FP64 and TF32 precision, respectively.

\subsection{Performance Breakdown}

\begin{figure}[t]
 \centering
 \includegraphics[width=0.96\linewidth]{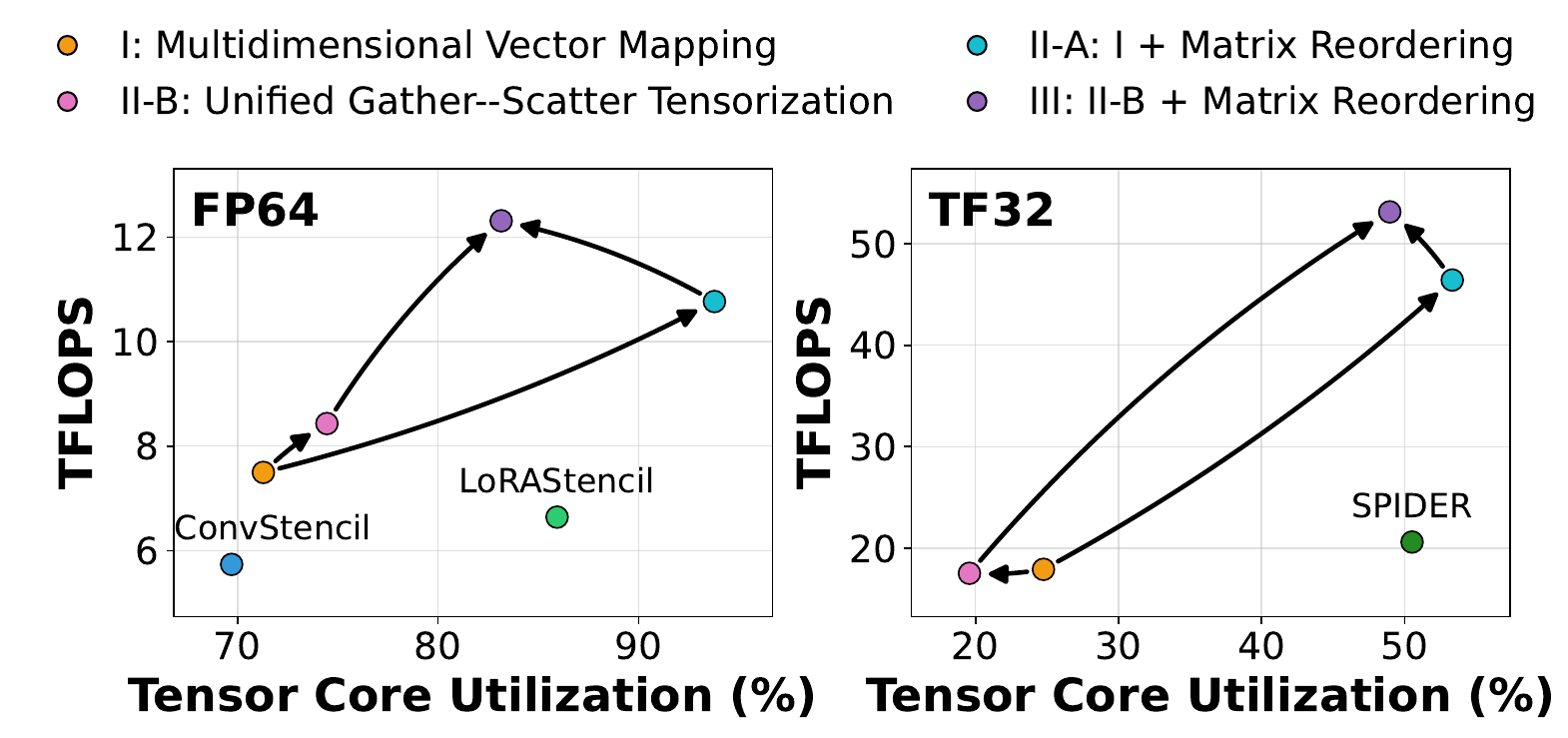}
\caption{Performance breakdown of PrismStencil for the j2d49pt kernel. Data points from other state-of-the-art approaches on the same kernel are included for reference.}
\label{fig:breakdown}
\end{figure}

Figure~\ref{fig:breakdown} presents a performance breakdown of PrismStencil under different combinations of computation and memory-access designs.

The results show that stencil tensorization must improve both computational efficiency and memory-access efficiency.
Reducing computational inflation increases TFLOPS per unit of Tensor Core utilization, but the achieved performance still depends on the utilization level.
For example, configuration II-B has lower computational inflation than configuration I.
Under TF32, however, both configurations are severely constrained by shared-memory traffic; consequently, II-B exhibits lower Tensor Core utilization while achieving nearly the same performance as I.
In contrast, improving memory access increases Tensor Core utilization and performance without changing the ratio between TFLOPS and Tensor Core utilization.
Once the memory bottleneck is alleviated, the performance benefit of reducing computational inflation becomes more pronounced.
Accordingly, the performance improvement from configuration II-A to configuration III is consistently greater than that from configuration I to configuration II-B.

\section{Discussion}

Sparse Tensor Cores exploit structured zeros to provide higher effective throughput.
Since stencil-to-matrix transformations naturally introduce zeros for nonneighboring grid-point pairs, the resulting weight matrices are well suited to sparse Tensor Core acceleration.

SPIDER~\cite{PPoPP26_Gu} extends the decomposition-based tensorization scheme characterized by Eq.~\eqref{eq:TCStencil-eff} to sparse Tensor Cores.
By setting $x=2r+2$ and $y=m$, it performs twice the intrinsic stencil operations before alignment.
Reordering the stencil-weight matrix to satisfy $2{:}4$ structured sparsity allows sparse Tensor Cores to skip half of these operations, leaving only alignment-induced computational inflation, given by $\text{align}(4r+2,k)\times \text{align}(2r+2,n)/((4r+2)\times (2r+2))$.
Although this factor is competitive with dense Tensor Core schemes under the same MMA shape, sparse Tensor Cores typically require larger shapes to reach peak throughput; for example, on NVIDIA A100 GPUs, FP16 dense and sparse Tensor Cores require $16\times8\times8$ and $16\times8\times32$, respectively.

As shown in Table~\ref{tab:TC-eff-cmp2}, SPIDER is highly sensitive to alignment.
At $r=7$, its matrix dimensions align well with the sparse MMA shape; although PrismStencil has a slightly higher computational inflation factor, both methods are already close to the lower bound of $1$, making the difference negligible.
At smaller radii, however, SPIDER incurs substantially greater alignment overhead and retains a higher computational inflation factor than PrismStencil even with kernel fusion~\cite{SC24_Zhang,PPopp24_Chen}.

For the same MMA shape, PrismStencil incurs lower computational inflation than existing dense Tensor Core tensorization schemes, making it better positioned to accommodate the larger MMA shapes required by sparse Tensor Cores while maintaining low computational overhead.
Extending PrismStencil to sparse Tensor Cores is therefore a promising direction and is left for future work.

\begin{table}[t]
\caption{Computational inflation factors of PrismStencil and SPIDER for two-dimensional box-shaped stencils with different radii, using the $16\times8\times8$ dense and $16\times8\times32$ sparse Tensor Core MMA shapes, respectively.}
\begin{center}
\begin{tabular}{*{5}{c}}
\toprule
Kernel Radius                   & 1 & 2 & 3 & 7\\
\midrule
SPIDER                          & 10.67  & 4.27    & 2.29    & 1.07\\
SPIDER $+$ Kernel Fusion          & 3.82   & 2.75    & 2.11    & 1.07\\
PrismStencil                    & 1.78   & 1.92    & 1.31    & 1.14\\
\bottomrule
\label{tab:TC-eff-cmp2}
\end{tabular}
\end{center}
\end{table}

\section{Related Work}
\label{sec:relatedwork}

Tiling~\cite{Malas17,Grosser14,Strzodka10,Strzodka11,Shrestha15,Bondhugula17} has been extensively studied to improve data reuse in stencil computations.
Representative techniques include tessellating~\cite{Yuan17,Yuan19}, split tiling~\cite{Henretty13,Grosser13}, diamond tiling~\cite{Bondhugula17,Bandishti12}, hexagonal tiling~\cite{Grosser14,Grosser18}, overlapped tiling~\cite{Krishnamoorthy07,Zhou12}, and N.5D blocking~\cite{SC10_Nguyen}.
N.5D blocking has become a mainstream tiling strategy for GPU stencil optimization, with representative systems including STENCILGEN~\cite{IEEE18_Rawat}, AN5D~\cite{CGO20_Matsumura}, and EBISU~\cite{ICS23_Zhang}.
Complementing these data-locality optimizations, vectorization~\cite{Henretty11,Yount15,Yuan21,Li21,Li22,Zhang25} has been widely used to improve the computational throughput of stencil kernels.

More recently, matrix acceleration units have enabled stencil computations to exploit higher-throughput matrix instructions on both CPUs and GPUs.
On CPUs, STOP~\cite{ICS24_Zhao} reformulates scatter-style stencil computations as sequences of vector outer products, while HStencil~\cite{SC25_Huang} jointly employs outer-product-based matrix units and vector units to accelerate stencil computation.
SMEStencil~\cite{TPDS2026_wang} improves memory-access efficiency for three-dimensional high-order stencils through architecture-specific optimizations.
On GPUs, Tensor Core-based approaches~\cite{PPopp24_Chen,ICS22_Liu,SC24_Zhang,SC25_Li,PPoPP26_Gu} reformulate stencil computations using dense or sparse matrix multiplications.
Dense approaches such as ConvStencil~\cite{PPopp24_Chen} and TCStencil~\cite{ICS22_Liu} incur substantial zero-padding overhead~\cite{gu2026needtensorcoresstencil}, whereas sparse approaches~\cite{SC25_Li,PPoPP26_Gu} reduce this overhead by exploiting structured sparsity.
FlashFFTStencil~\cite{PPopp25_Haozhi} instead adopts an FFT-based formulation, which can reduce computational complexity in suitable cases.

Other directions of stencil optimization include composite stencil optimization that combines DAG-based transformations with other optimization techniques~\cite{IPDPS19_Rawat,SC24_Liu} and distributed-memory stencil optimization that improves scalability and communication efficiency~\cite{CCGRID18_Ying,SC20_Wang,IPDPS20_Yu,PC22_Johannes}.

\section{Conclusion}

This work presented PrismStencil, a stencil tensorization algorithm that addresses the computational and memory-access inefficiencies of existing Tensor Core stencil methods.
PrismStencil introduces multidimensional vector mapping and a unified gather--scatter tensorization scheme to better capture the multidimensional structure of stencil computations and reduce the redundant computation introduced by tensorization.
It further develops a cross-product data-reuse mechanism that exploits the inherent overlap among neighboring stencil updates to enhance data reuse.
Across representative stencil kernels on an NVIDIA A100 GPU, PrismStencil achieves average speedups of $2.13\times$ and $2.17\times$ over state-of-the-art Tensor Core stencil methods with FP64 and TF32 precision, respectively.


\bibliographystyle{ACM-Reference-Format}
\bibliography{sample-base}

\appendix

\section{Problem Size}

\begin{table*}[htbp]
\caption{Problem size configuration for FP64 on NVIDIA A100}
\begin{center}
\begin{tabular}{*{6}{c}}
\toprule
Kernel       & j2d5pt               & j2d9pt               & j2d25pt              & j2d49pt              & j2d81pt              \\
\midrule
ConvStencil  & $1000\times 9216^2$  & $1000\times 9216^2$  & N/A                   & $1000\times 9216^2$  & N/A                   \\
LoRAStencil  & $1000\times 9216^2$  & $1000\times 9216^2$  & N/A                   & $1000\times 9216^2$  & N/A                   \\
AN5D         & $1000\times 8192^2$  & $1000\times 8192^2$  & $1000\times 8192^2$  & $1000\times 8192^2$  & $1000\times 8192^2$  \\
EBISU        & $1000\times 8352^2$  & $1000\times 8784^2$  & $1000\times 8640^2$  & $1000\times 8784^2$  & N/A                   \\
PrismStencil & $1000\times 9648^2$  & $1000\times 9648^2$  & $1000\times 8352^2$  & $1000\times 9360^2$  & $1000\times 7488^2$  \\
\bottomrule
\label{tab:Problem Size A100-fp64}
\end{tabular}
\begin{tabular}{*{4}{c}}
\toprule
Kernel       & j3d7pt                                      & j3d27pt                                     & j3d125pt                                    \\
\midrule
ConvStencil  & $1000\times 2560\times 288\times 384$     & $1000\times 2560\times 288\times 384$     & N/A                                         \\
LoRAStencil  & $1000\times 640\times 576\times 576$      & $1000\times 1280\times 480\times 576$     & N/A                                         \\
AN5D         & $1000\times 640^3$                         & $1000\times 640^3$                         & $1000\times 640^3$                         \\
EBISU        & $1000\times 2560\times 288\times 384$     & $1000\times 2560\times 288\times 384$     & N/A                                         \\
PrismStencil & $1000\times 2560\times 270\times 368$     & $1000\times 2560\times 270\times 368$     & $1000\times 2560\times 288\times 368$     \\
\bottomrule
\end{tabular}
\end{center}
\end{table*}

\begin{table*}[htbp]
\caption{Problem size configuration for FP32/TF32 on NVIDIA A100}
\begin{center}
\begin{tabular}{*{6}{c}}
\toprule
Kernel       & j2d5pt               & j2d9pt               & j2d25pt               & j2d49pt              & j2d81pt              \\
\midrule
TCStencil    & $1000\times 8192^2$  & $1000\times 8192^2$  & $1000\times 8192^2$   & N/A                  & N/A                  \\
SPIDER       & $1000\times 8192^2$  & $1000\times 8192^2$  & $1000\times 8192^2$   & $1000\times 8192^2$ & $1000\times 8192^2$ \\
AN5D         & $1000\times 8192^2$  & $1000\times 8192^2$  & $1000\times 8192^2$   & $1000\times 8192^2$ & $1000\times 8192^2$ \\
EBISU        & $1000\times 8568^2$  & $1000\times 8640^2$  & $1000\times 8352^2$   & $1000\times 9000^2$ & N/A                  \\
PrismStencil & $1000\times 7560^2$  & $1000\times 7560^2$  & $1000\times 10368^2$  & $1000\times 8256^2$ & $1000\times 9216^2$ \\
\bottomrule
\label{tab:Problem Size A100-fp32}
\end{tabular}
\begin{tabular}{*{4}{c}}
\toprule
Kernel       & j3d7pt                                      & j3d27pt                                     & j3d125pt                                    \\
\midrule
AN5D         & $1000\times 640^3$                         & $1000\times 640^3$                         & $1000\times 640^3$                         \\
EBISU        & $1000\times 2560\times 288\times 384$     & $1000\times 2560\times 288\times 384$     & N/A                                         \\
PrismStencil & $1000\times 2560\times 252\times 360$     & $1000\times 2560\times 252\times 360$     & $1000\times 2560\times 252\times 352$     \\
\bottomrule
\multicolumn{4}{l}{$^{\mathrm{a}}$ TCStencil and SPIDER are omitted for 3D kernels because their AD/AE artifacts do not provide}\\
\multicolumn{4}{l}{implementations for 3D stencils.}
\end{tabular}
\end{center}
\end{table*}

\end{document}